\documentclass{aa}  

\usepackage{graphicx}
\usepackage{txfonts}
\begin{document}

%   \title{Geophysical Assessment of Habitability for the TRAPPIST-1 Exoplanets}
    \title{Tidal Heating and the Habitability of the TRAPPIST-1 Exoplanets}

%   \subtitle{I. Overviewing the $\kappa$-mechanism}

   \author{Vera Dobos
          \inst{1,2,3,4}\thanks{E-mail: dobos@konkoly.hu}
          Amy C. Barr\inst{5}
          \and
          L\'aszl\'o L. Kiss\inst{1,6}
          }

   \institute{Konkoly Observatory, Research Centre for Astronomy and Earth Sciences, \\Hungarian Academy of Sciences, H--1121 Konkoly Thege Mikl\'os \'ut 15-17, Budapest, Hungary
   \and Geodetic and Geophysical Institute, Research Centre for Astronomy and Earth Sciences, \\Hungarian Academy of Sciences, H--9400 Csatkai Endre u. 6-8., Sopron, Hungary
   \and Hungarian E\"otv\"os Fellow, Tempus Public Foundation, H--1077 K\'ethly Anna t\'er 1., Budapest, Hungary
   \and ELTE E\"otv\"os Lor\'and University, Gothard Astrophysical Observatory, Szombathely, Szent Imre h. u. 112, Hungary
   \and Planetary Science Institute, 1700 E. Ft. Lowell, Suite 106, Tucson, AZ 85719, USA
    \and Sydney Institute for Astronomy, School of Physics A28, University of Sydney, NSW 2006, Australia
             }

   \date{Received September 17, 2018; accepted February 11, 2019}

% \abstract{}{}{}{}{} 
% 5 {} token are mandatory
 
  \abstract
   {New estimates of the masses and radii of the seven planets orbiting the ultracool M-dwarf TRAPPIST-1 star permit improved modelling of their compositions, heating by tidal dissipation, and removal of tidal heat by solid-state convection.}
   {Here, we compute the heat flux due to insolation and tidal heating for the inner four planets.}
   {We apply a Maxwell viscoelastic rheology to compute the tidal response of the planets using the volume-weighted average of the viscosities and rigidities of the metal, rock, high-pressure ice and liquid water/ice I layers.}
   {We show that TRAPPIST-1d and e can avoid entering a runaway greenhouse state. Planet e is the most likely to support a habitable environment, with Earth-like surface temperatures and possibly liquid water oceans.
Planet d also avoids a runaway greenhouse, if its surface reflectance is at least as high as that of the Earth.  
Planets b and c, closer to the star, have heat fluxes high enough to trigger a runaway greenhouse and support volcanism on the surfaces of their rock layers, rendering them too warm for life.  
Planets f, g, and h are too far from the star to experience significant tidal heating, and likely have solid ice surfaces with possible subsurface liquid water oceans.}
   {}


   \maketitle
%
%________________________________________________________________

\section{Introduction}

The ultracool M-dwarf star TRAPPIST-1 appears to harbour seven roughly Earth-sized planets, among which
planets d, e, f, and g may be potentially habitable \citep{gillon17,papaloizou17,unterborn18,barr18}.
Prior works have addressed habitability in terms of stellar activity, water loss, and implications of tidal locking
\citep{vida17,garraffo17,bourrier17,checlair17,dong18, turbet18}, but none have addressed the role of the planets'
interior geodynamics in maintaining a habitable environment.
Planets can be habitable only if they avoid entering a ``runaway greenhouse'' state, in which a high geothermal and/or solar heat flux causes all of their
surface water to evaporate irreversibly into a thick atmosphere \citep{kasting93}.

Planets that experience a runaway greenhouse are generally not considered to be habitable, because their water has evaporated and been split into hydrogen and oxygen in the planet's upper atmosphere by photodissociation \citep{kasting93}.  The hydrogen molecules can easily escape to space, and as a consequence, water molecules cannot re-form, even if the planet's temperature later decreases.  A planet may enter a runaway greenhouse if its net heat flux, $F_\mathrm{glob}$, exceeds a critical energy flux, $F_\mathrm{RG}$, at the top of a water-rich atmosphere.  
%Table \ref{table:params} summarises the latest estimates of the masses and radii of the planets \citep{delrez18, grimm18} and Table \ref{table:results} lists critical values of $F_{RG}$ for planets b through e, calculated using the planet's sizes and masses and the optical/physical properties of water (see Methods) \citep{pierrehumbert10}. 
To determine if a runaway greenhouse state is likely for any of the TRAPPIST-1 planets, we compare $F_\mathrm{RG}$ to the global heat flux from the planet $F_\mathrm{glob}=F_{\star}+F_\mathrm{int}$, where $F_{\star}$ is the energy received from sunlight, and $F_\mathrm{int}$ is the heat flux from the planet's interior. 

The TRAPPIST-1 planets are heated from within by tidal dissipation \citep{barr18}, which is the dominant contribution to $F_\mathrm{int}$.  To determine $F_\mathrm{int}$, we equate the heating rate from tidal dissipation to the heat flux from solid-state convection \citep{barr18}, the dominant mode of removing heat from the deep interior of a solid planet \citep{SM2000}.  The planets orbit close to the parent star in eccentric orbits, and experience a time-variable tidal gravitational force from the star.  Tidal forces raise and lower a bulge on each planet; as the height of the bulge changes, the planet does work against its own internal rigidity, resulting in the dissipation of orbital energy in the planet's interior \citep{peale78}.  This phenomenon is responsible for the prodigious volcanism on Jupiter's moon Io \citep{peale79} and cryovolcanism on Saturn's moon Enceladus \citep{porco06}.  

Applying N-body simulations to planets of the TRAPPIST-1 system, \citet{luger17} showed that due to tidal damping, the orbital eccentricities quickly decrease to values lower than 0.01, but the orbits do not circularize because of planet--planet interactions. The resonance chain maintains the eccentricities of the TRAPPIST-1 planets, similarly to the case of the three Galilean moons in the Solar System (Io--Europa--Ganymede). The non-circularizing orbits have a key role in maintaining tidal heat generation for long timescales.

The tidal heating rate for each planet depends strongly on its composition and temperature \citep{segatz88}.  Ice, rock, and iron, the most likely planetary materials \citep{gillon17,unterborn18,grimm18}, each behave differently in response to time-variable tidal stresses; the response of each material is also sensitively dependent on its temperature.  The temperature in the planet's interior is controlled by the balance between tidal heat generation and heat transport \citep{barr18, henning09, dobos15}.   

%__________________________________________________________________

\section{Methods}

\begin{figure}
\centerline{\includegraphics[width=16pc]{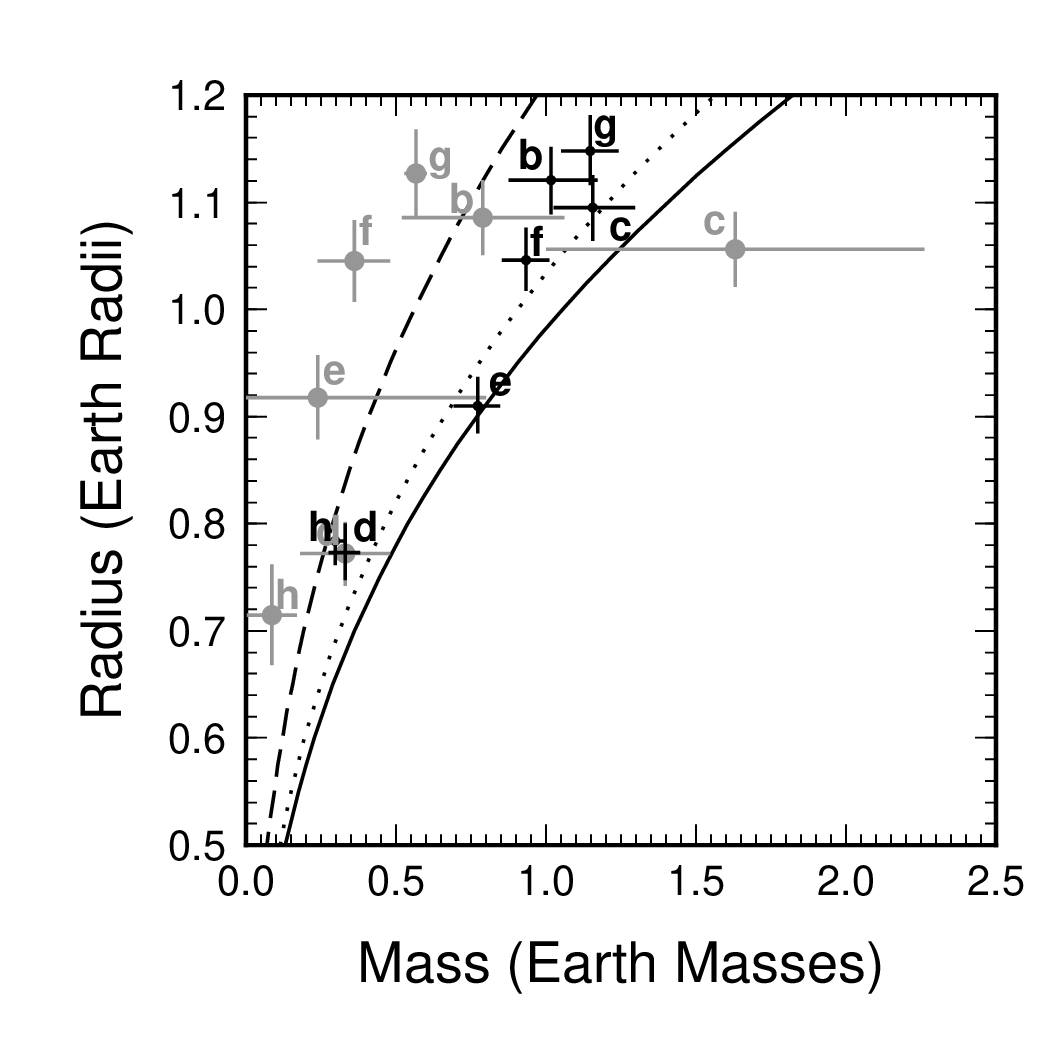}}
\caption{\label{fig:mass-radius} The most recent planetary masses and radii \citep{grimm18, delrez18} (black), compared to prior estimates from Wang et al., \citep{wang17} (gray). Most of the planets could contain ice, rock and metal, with 75\% ice by mass (dashed, with 12.5\% rock and 12.5\% iron), pure rock (dotted) or 50\% rock by mass, with 25\% iron, and 25\% ice (solid curve).}
\end{figure}

To evaluate $F_\mathrm{int}$, we compute the tidal response of each planet assuming it is composed of a mixture of H$_2$O, metal, and rock, using a Maxwell viscoelastic rheology characterized by a single uniform viscosity and rigidity for the planet given by the volume-weighted average of the viscosities and rigidities of each material \citep{barr18, henning09}.  The tidal heating rate is set equal to the heat flux from solid-state convection across the planet's rock mantle \citep{barr18,SM2000}. We consider tidal heating in the planets' rock mantles to be the dominant source of heat in the planets at this time, and solid-state convection as the means by which this heat is transported.  We do not include other heat sources such as radiogenic heating, or residual heat from accretion or differentiation.  If the abundances of heat-producing elements in the TRAPPIST-1 planets are similar to those in the planets in our solar system, tidal heating will be much larger than radiogenic heating \citep{Frank2014}, especially considering the age of the system ($\sim$8~Gyr) \citep{burgasser2017}. We search for the rock mantle temperature ($T_\mathrm{eq}$) at which the tidal heating rate is equal to the convective heating rate.  The tidal heat flux evaluated at $T_\mathrm{eq}$ is equal to $F_\mathrm{int}$.  This approach reproduces the spacecraft-measured value of tidal heat flux and estimated interior temperatures for Jupiter's moon Io to better than 20\% \citep{barr18}.  Because the masses and radii of the planets still have uncertainties at the $\sim$10\% level \citep{grimm18, delrez18}, a variety of bulk compositions and interior structures are possible (see Fig \ref{fig:mass-radius}).  We explore a range of interior structures and compositions assuming each planet is composed of uniform-density (incompressible) iron, compressed Bridgmanite (MgSiO$_3$) as a proxy for silicate rock, high-pressure ice polymorphs, and ice I/liquid water \citep{barr18}.  

\begin{table}
\begin{center}
\begin{tabular}{lcccc}
\hline
Planet  & $a$ (AU) & $e$ $(10^{-3})$   &  $M$ ($M_{\oplus}$) & $R$ ($R_{\oplus}$) \\
\hline
b & 0.01154775 & $6.22 \pm 3.04$ & 1.017$^{+0.154}_{-0.143}$ & 1.121$^{+0.031}_{-0.032}$ \\
c & 0.01581512 & $6.54 \pm 1.88$ & 1.156$^{+0.142}_{-0.131}$ & 1.095$^{+0.030}_{-0.031}$ \\
d & 0.02228038 & $8.37 \pm 0.93$ & 0.297$^{+0.039}_{-0.035}$ & 0.784$\pm 0.023$  \\
e & 0.02928285 & $5.10 \pm 0.58$ & 0.772$^{+0.079}_{-0.075}$ & 0.910$^{+0.026}_{-0.027}$ \\
f &  0.03853361 & $10.07 \pm 0.68$ & 0.934$^{+0.080}_{-0.078}$ & 1.046$^{+0.029}_{-0.030}$ \\
g & 0.04687692 & $2.08 \pm 0.58$ & 1.148$^{+0.098}_{-0.095}$ & 1.148$^{+0.032}_{-0.033}$ \\
h & 0.06193488 & $5.67 \pm 1.21$ & 0.331$^{+0.056}_{-0.049}$ & 0.773$^{+0.026}_{-0.027}$ \\
\hline
\end{tabular} 
\caption{Planetary semi-major axes ($a$), eccentricities ($e$), masses ($M$, scaled by the mass of the Earth $M_{\oplus}=5.98\times10^{24}$ kg), and radii ($R$, scaled by Earth's radius $R_{\oplus}=6371$ km) from Grimm et al., \citep{grimm18} and Delrez et al. \citep{delrez18}.  \label{table:params} } 
\end{center}
\end{table}

Planetary parameters are summarised in Table \ref{table:params}.  The critical heat flux for a runaway greenhouse, $F_\mathrm{RG}$ is calculated using the formulation of \citet{pierrehumbert10} \citep[see also ][]{barr18}. 
The heat flux from stellar radiation, 
\begin{equation}
   \label{eq:Fstar}
      F_\star = \frac { L_\star \left( 1 - A_\mathrm{B} \right) } { 16 \pi a^2 \sqrt{1 - e^2} } \, ,
\end{equation}
where $L_\star=5.24\times10^{-4}L_{\odot}$ is the stellar luminosity, $A_\mathrm{B}$ is the Bond albedo of the planet, $a$ and $e$ are the semi-major axis and the eccentricity of the planet's orbit, respectively. In the calculations the uncertainties in the semi-major axis (a) and eccentricity (e) are not taken into account. Greenhouse and cloud feedbacks were neglected in the calculation of $F_{\star}$. We use three Bond albedo values for each planet: 0.1, 0.3 (an Earth-like value) and 0.5.

We use PREM (Preliminary Earth Reference Model) values for the densities of iron ($\rho_\mathrm{i}=12,000$ kg/m$^3$) and rock ($\rho_\mathrm{r}=5,000$ kg/m$^3$) \citep{dziewonski81}, a representative density for the compressed ice phases (high-pressure polymorphs, or hpp) $\rho_\mathrm{hpp}$=1300 kg/m$^3$ \citep{hobbs74}, and $\rho_\mathrm{iw}=1000$ kg/m$^3$ for ice I and liquid water. For each possible planetary mass and radius, we compute a suite of values for the volume fractions ($\phi$) of ice~I/water, hpp ice, rock, and metal that satisfy:
\begin{eqnarray}
\phi_\mathrm{iw}+\phi_\mathrm{hpp}+\phi_\mathrm{r}+\phi_\mathrm{Fe}&=&1, \label{eq:1}\\
\phi_\mathrm{iw}\rho_\mathrm{iw}+\phi_\mathrm{hpp}\rho_\mathrm{hpp}+\phi_\mathrm{r}\rho_\mathrm{r}+\phi_\mathrm{fe}\rho_\mathrm{Fe} &=&\rho, \label{eq:2}
\end{eqnarray}
where $\rho$ is the mean density of the planet.  If $\rho<\rho_\mathrm{r}$, the planet must contain H$_2$O in some form.  For each water-bearing structure, $\phi_\mathrm{iw}$ is estimated by determining the thickness of the low-density ice~I (or possibly water) layer at the surface, $z_{209}\approx(209 \textrm{ MPa})/(\rho_\mathrm{iw}g)$, where 209 MPa is the pressure at which water undergoes the phase transition to the high-pressure ice polymorphs, and $g=GM_\mathrm{pl}/R_\mathrm{pl}^2$.  The planetary radius at which the phase transition occurs, $R_\mathrm{iw}\approx R_\mathrm{pl}-z_{209}$, which gives $\phi_\mathrm{iw}=1-(R_\mathrm{iw}/R_\mathrm{pl})^3$.
We explore a range of $\phi$ values for each of the remaining components to determine which values satisfy eq. (\ref{eq:1}) and~(\ref{eq:2}).

The tidal heat flux $F_\mathrm{tidal}$ is calculated using a viscoelastic model for a homogeneous body with a strongly temperature-dependent viscosity $\eta$ and shear modulus $\mu$, constructed to mimic the effect of a layered planet \citep{barr18},
\begin{eqnarray}
\eta(T)&=&\phi_\mathrm{iw}\eta_\mathrm{iw}(T) + \phi_\mathrm{hpp}\eta_\mathrm{hpp}(T)+\phi_\mathrm{r}\eta_\mathrm{r}(T),\\
\mu(T) &=& \phi_\mathrm{iw}\mu_\mathrm{iw}(T) + \phi_\mathrm{hpp}\mu_\mathrm{hpp}(T)+\phi_\mathrm{r}\mu_\mathrm{r}(T).
\end{eqnarray}
Governing parameters for the rheologies of each material can be found in \citet{barr18}. The viscosity and shear modulus for rock are calculated assuming a dunite composition \citep{Berckhemer82,henning09,karato95,SM2000}.  The viscosity of rock is calculated under the assumption that the rock is dry and deforms under diffusion creep \citep{karato95, SM2000}.  The viscosity and rigidity of each material depends strongly on temperature, and both include the effect of melt.  As the melt fraction increases, the viscosity and rigidity of each material decreases, giving a concomitant decrease in the amount of tidal heating \citep{henning09}. In our model, in each planet, tidal heating is maximized at the temperature at which the ratio between the viscosity and rigidity, the Maxwell time, is equal to the orbital period of the planet: $\eta (T)/\mu (T) \approx T_{orbit}$ \citep{segatz88}.  For TRAPPIST-1b through e, this occurs at $T_{eq} \sim 1600$ K, for the viscosity and rigidity parameters used here and in our prior study \citep{barr18}.  Of course, this can depend on the choice of viscosity and rigidity parameters ((e.g., dry vs. wet rheologies), as well as the choice of rheological model (e.g., Maxwell, standard linear solid, Andrade, etc.), see \citet{henning09} for a full discussion of these effects. We neglect the viscosity and rigidity of iron because tidal deformation of the planets' iron cores will be negligible due to the restoring force imposed by the rock and ice mantle that lies atop the core in each body \citep{henning09}. Tidal dissipation in the iron portion of the Earth is a small contributor to our planet's global heat flux, providing only 1\% of the total global heat flux \citep{lay08}.  The tidal heat flux \citep{segatz88}
\begin{equation}
F_\mathrm{tidal} = \frac{21}{8 \pi R^2} Im(k_2^*) \frac{R^5\omega^5 e^2 }{G},
\end{equation}
with orbital frequency $\omega=2\pi/P$, orbital period $P$, and eccentricity $e$, and $k_2^*$ \citep{segatz88}
\begin{equation}
k_2^*=\frac{(3/2)}{1+\frac{19 \mu^*}{2 \rho g R}}.
\end{equation}
The planet's resistance to tidal deformation is described by a complex shear modulus $\mu^*=M_1+iM_2$ where $i=\sqrt{-1}$, $M_1=\mu(T)(\omega \eta(T))^2/C$, $M_2=\mu(T)^2 (\omega \eta(T))/C$, and $C=\mu{T}^2 + (\omega \eta(T))^2$ \citep{peltier74}.  The tidal heat generated inside the planet is transported to the surface by solid-state convection \citep{SM2000, barr18}.  The convective heat flux is calculated using scaling relationships for stagnant lid convection in internally heated rock mantles \citep{SM2000, barr18}, 
\begin{equation}
F_\mathrm{int}=0.53 \bigg(\frac{Q^*}{R_G T_\mathrm{eq}^2}\bigg)^{-4/3} \bigg(\frac{\rho_r g \alpha k_\mathrm{therm}^3}{\kappa_\mathrm{therm} \eta_r(T_\mathrm{eq})}\bigg)^{1/3}, \label{eq:heatflux}
\end{equation}
with gas constant $R_G=8.314$ J/mol K, planetary surface gravity $g=GM/R$, activation energy for volume diffusion in rock $Q^*=300$ kJ/mol, coefficient of thermal expansion $\alpha=3 \times 10^{-5}$ K$^{-1}$, rock thermal conductivity $k_\mathrm{therm}=3.2$ W/m/K, thermal diffusivity $\kappa_{therm}=k_{therm}/\rho_r C_p$ and rock specific heat $C_p=1200$ J/kg \citep{SM2000}, and rock viscosity $\eta_r$ \citep{barr18}.  %We assume stagnant lid convection is occurring in the rock mantles of the TRAPPIST-1 planets because it is not known whether ``plate tectonics'' or widespread advection of melt similar to the process that drives volcanism on Io might be possible.  The factors that cause a given rocky planet to be losing heat in ``stagnant-lid mode'' versus ``plate tectonics'' mode not entirely well-understood, even for planets in our own solar system \citep{bercovici03,bercovici14}; thus, it is still premature to speculate about more exotic modes of heat loss on the TRAPPIST-1 planets.

Uncertainties in the geothermal and the critical runaway greenhouse heat fluxes arise from uncertainties in the planetary masses and radii.  Uncertainties in the values of tidal and global heat fluxes additionally arise from different possible values of albedo and orbital eccentricity. The uncertainties quoted here for our $T_\mathrm{eq}$ arise from the different possible interior structures, namely the uncertainties in the volume fractions of different materials.  A further source of uncertainty is the rheology appropriate for rock: if the rock in the planets' interiors is hydrated, the viscosity and rigidity can be altered, which can affect the value of $T_{eq}$.  Comparing to previous findings \citep{barr18}, all planets except for planet d have weaker tidal heating as a result of new constraints on their physical and orbital parameters.

\section{Results}

\begin{figure*}
\centerline{\includegraphics[width=26pc]{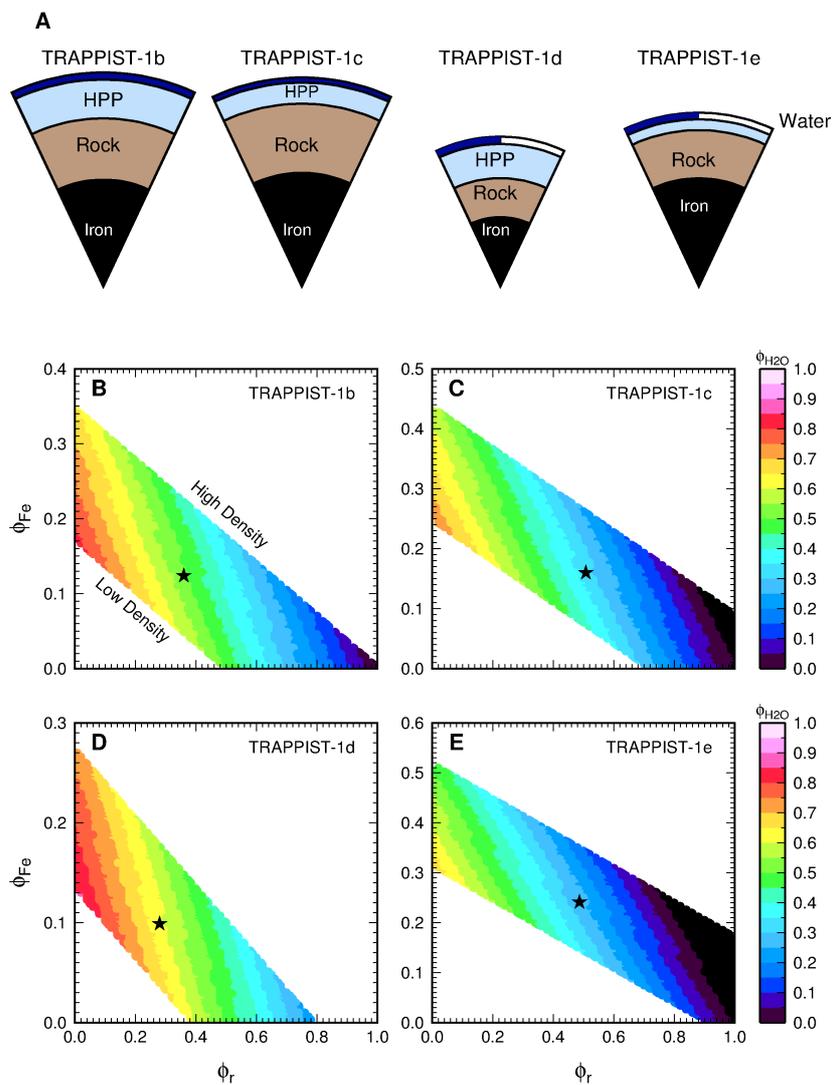}}
\caption{\label{fig:colors} Compositions and interior structures
of planets TRAPPIST-1b through e.  (A) Representative possible interior structures of planets b through e, with layers of liquid water (dark blue), high-pressure ice polymorphs (light blue), rock (brown), and iron (black).  Planets d and e are far enough from the TRAPPIST-1 star to potentially have a layer of solid ice (white) on their surfaces.  (B) -- (E) Compositions of planets b through e, where colours indicate the volume fraction of H$_2$O permitted in each planet ($\phi_\mathrm{H2O}=\phi_\mathrm{iw}+\phi_\mathrm{hpp}$) as a function of its volume fraction of iron ($\phi_\mathrm{Fe}$) and rock ($\phi_\mathrm{r}$).  Black stars indicate the cases shown in panel A.  Planets b and d are substantially more H$_2$O-rich than planets c and e. }
\end{figure*}

\begin{figure*}
\centerline{\includegraphics[width=36pc]{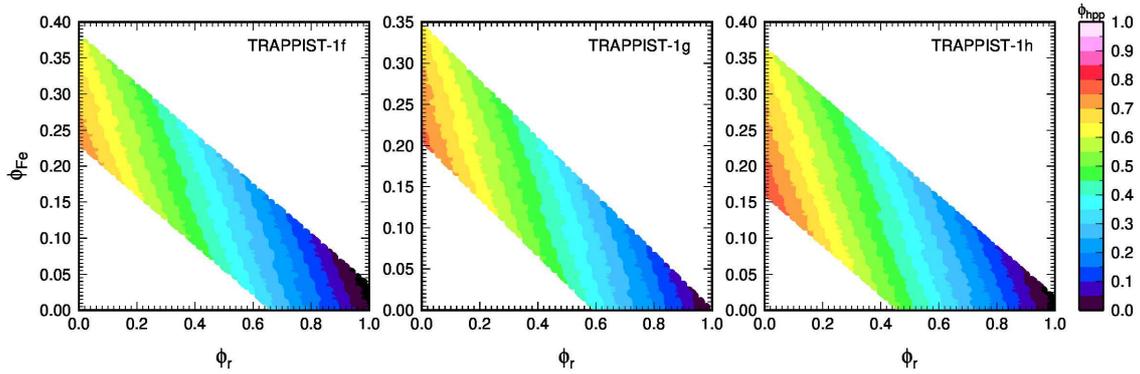}}
\caption{Same as panels B through E of Figure \ref{fig:colors}, but for TRAPPIST-1f, g, and h.  Colours indicate the volume fraction of H$_2$O permitted in each planet ($\phi_\mathrm{H2O}$) as a function of the volume fraction of iron ($\phi_\mathrm{Fe}$) and rock ($\phi_\mathrm{r}$).  \label{fig:colors-outer}}
\end{figure*}

Figure \ref{fig:colors} illustrates the values of the volume fraction, $\phi$ that are possible for each material, for planets b through e, as well as fiducial interior structures for each body.  (Similarly, Fig. \ref{fig:colors-outer} shows the possible interior structures for the outer three planets.)  The width of the coloured region originates from the uncertainties in the mass and radius estimates of the planets; the top-right (bottom-left) part of this area includes the high-density (low-density) cases for each planet, as indicated in panel B. The vast majority of valid interior structures for all the planets include significant H$_2$O. Despite its proximity to the star, planet b must contain at least 0.2\% H$_2$O by volume, consistent with prior predictions \citep{unterborn18, barr18, wang17, grimm18}, and we further find the vast majority of the possible structures for planet b have substantial quantities of water ($\sim 20-70$\%). The remaining planets are also probably water-rich, most notably planet d, where a volume fraction of H$_2$O$\gtrsim$0.25 is required by the mass and radius data.

\begin{table}
\begin{center}
\begin{tabular}{lcccc}
\hline
Planet  &  $F_\mathrm{RG}$ & $T_\mathrm{eq}$ (K) & $F_\mathrm{int}$ & $F_\mathrm{glob}$  \\
\hline
b &$289.1^{+5.8}_{-7.0}$ &$1683^{+28}_{-83}$ &$1.02^{+0.95}_{-0.93}$ &$937.1^{+268.4}_{-268.4}$\\
c &$294.5^{+5.4}_{-5.9}$ &$1659^{+22}_{-59}$ &$0.62^{+0.42}_{-0.53}$ &$499.7^{+143.0}_{-143.1}$\\
d &$273.2^{+5.0}_{-5.8}$ &$1645^{+17}_{-45}$ &$0.26^{+0.14}_{-0.21}$ &$251.7^{+72.0}_{-72.1}$\\
e &$292.5^{+5.7}_{-4.5}$ &$1604^{+14}_{-4}$ &$0.14^{+0.04}_{-0.14}$ &$145.7^{+41.6}_{-41.7}$\\
\hline
\end{tabular} 
\caption{\label{table:results} Equilibrium mantle temperature ($T_\mathrm{eq}$ in Kelvin), interior heat flux ($F_\mathrm{int}$), global heat flux 
($F_\mathrm{glob}$), and runaway greenhouse limit ($F_\mathrm{RG}$).  All heat fluxes have units of W/m$^2$.} 
\end{center}
\end{table}

Table \ref{table:results} summarises the values of $T_\mathrm{eq}$, $F_\mathrm{tidal}$, $F_\mathrm{RG}$, and $F_\mathrm{glob}$ for planets b through e. Planets d and e have a total heat flux from tidal heating and insolation that is below the critical heat flux for a runaway greenhouse, and could be potentially habitable from a geophysical point of view.  Planet d avoids a runaway greenhouse only if its Bond albedo is greater or approximately equal to 0.3, indicating a surface at least as reflective as the surface of the Earth, which could be possible if the planet has a thick atmosphere with highly reflective clouds \citep{turbet18}.  In this case, planet d could have liquid water at its surface at the substellar point provided it is tidally locked in a 1:1 spin--orbit resonance \citep{turbet18}.  It is not clear whether TRAPPIST-1d is synchronously rotating because perturbations from the other planets may prevent tidal locking which result in a larger illuminated area where surface ice can melt \citep{vinson17}.  Since the planet is close to the runaway greenhouse state due to the received stellar irradiation, the moderate tidal heating rate that is present in the body might transform it into a ``tidal Venus'' \citep{barnes13}.  We find that planet e avoids a runaway greenhouse regardless of its albedo, bolstering the already strong case for its habitability based solely on climate modeling \citep{wolf17, turbet18}.  Planet e can maintain global surface oceans through greenhouse warming from H$_2$O alone \citep{turbet18}; alternatively, Earth-like temperatures can be maintained with an atmosphere of 1 bar of N$_2$ and 0.4 bar CO$_2$, or a pure CO$_2$ atmosphere with a pressure of 1.3 bar \citep{wolf17}. 

If planets d or e were ice-covered, they could still sustain biospheres fueled by the planet's interior tidal dissipation, for example, clustered around hydrothermal vents at the top of the internally heated rock layer, as has been suggested for Jupiter's moon Europa \citep{vance07}. Since both planets have equilibrium rock mantle temperatures greater than the melting point of silicate, heat dissipation from the rock layer can locally melt the high-pressure ice to provide a warmer environment where the rock interacts with water through the vents and might support simple chemoautolithotropic life \citep{vance07}. The surface ice layer would shield the biosphere from the strong stellar flare activity of the TRAPPIST-1 star \citep{vida17, garraffo17}, but would serve as a barrier to detection of the biosphere through telescopic observations.  Hydrothermal activity would be more pronounced on planet d, whose equilibrium temperatures can be as high as 1660 K, indicating the mantle could have as much as 35\% melt by volume \citep{mckenzie88}, similar to Io's asthenosphere \citep{barr18}.  The volume fraction of melt in the rock mantle of planet e would be more modest, less than 10\% \citep{mckenzie88}, which might not lend itself to efficient extraction by volcanism.

Planets b and c have heat flows far in excess of that required to trigger a runaway greenhouse.  The apparent H$_2$O-rich composition of planet b seems to be at odds with a thick atmosphere given how quickly atmospheres may be stripped from the planets due to their proximity to the star \citep{wolf17}.  One explanation is that the planets only recently migrated to their present orbital configuration \citep{unterborn18}.  The heat flows on both bodies are similar to that observed on Io ($\sim$1--2 W/m2) \citep{Spencer00, Veeder04}, and correspondingly, the planets have high mantle temperatures suggestive of melt-rich asthenospheres that could drive extensive volcanic activity \citep{barr18}.

According to our model, planets f, g, and h are too far from the TRAPPIST-1 star to experience significant tidal effects from the star and are likely covered with thick ice layers at their surfaces \citep{barr18}.  The presence of an atmosphere does not modify these conclusions: any atmosphere on these bodies would quickly condense on their cold surfaces \citep{wolf17}.  Planet--planet interactions (which were not considered in this work) and the existing mean motion resonances between the planets, however, could contribute to the tidal heating rates considerably, especially in planet g, assuming that the planets are tidally locked to the star \citep{wright18, lingam18}. Given that every large body in our Solar System with a thick ice mantle also harbours a liquid water ocean \citep{hussmann06}, it seems likely that the planets, even with ice surfaces, would have liquid water oceans beneath kilometers of ice. Our model considers tidal heating to occur mainly in the rock mantles of the planets; a layered tidal model that partitions dissipation in each layer of the planet based on its rheology could shed more light on possible tidal heating in the outer three planets in the system \citep[e.g., ][]{roberts08}. Better constrained radius and mass measurements are also critical for the characterization of planetary interiors since the H$_2$O content of the bodies greatly depend on these estimates \citep{dorn18}.

\begin{acknowledgements}
We thank the referee for useful questions and comments. V. D. thanks the Max Planck Institute for Solar System Research for providing a work environment. V. D. is supported by the Hungarian National Research, Development, and Innovation Office (NKFIH) grants K-119993, K-115709, and GINOP-2.3.2-15-2016-00003.  A. C. B. acknowledges support from NASA Habitable Worlds 80NSSC18K0136.  This work has been partially funded by the Tempus Public Foundation, project number 156411.
\end{acknowledgements}

%-------------------------------------------------------------------

\bibliographystyle{aa}
\bibliography{ref}

\end{document}